\documentclass[12pt]{article}
\pdfoutput=1
\usepackage{graphicx}
\usepackage{epstopdf}
\usepackage{amsmath}
\usepackage{amsfonts}
\usepackage{amssymb}
\usepackage{color}
\usepackage{mathrsfs}
\usepackage{graphicx} 
\usepackage{bm} 
\usepackage{epsfig}
\usepackage{amsmath, amssymb}
\usepackage{hyperref}
\usepackage{color}
\usepackage{colordvi}
\usepackage{comment}
\usepackage{graphicx}

\newcommand{\trac}[2]{{\displaystyle\frac{#1}{#2}}}

\newcommand\ee{\end{equation}}
\newcommand\be{\begin{equation}}
\newcommand\eea{\end{eqnarray}}
\newcommand\bea{\begin{eqnarray}}

\newcommand\ket[1]{{| {#1} \rangle}}
\newcommand\bra[1]{{\langle {#1} |}}
\newcommand\vev[1]{{\langle {#1} \rangle}}
\renewcommand\({\left(}
\renewcommand\){\right)}
\newcommand\ex[1]{{\mbox{e}^{#1}}}
\newcommand\eq[1]{Eq.~(\ref{#1})}

\renewcommand{\d}{{\rm{d}}}

\newcommand{\del}{\partial}
\def\fnl{f_{\rm NL}}

\newcommand\gev{\,\mbox{GeV}}

\newcommand\ie{{\it i.e.}~}
\newcommand\eg{{\it e.g.}~}

\renewcommand\({\left(}
\renewcommand\){\right)}
\renewcommand\[{\left[}
\renewcommand\]{\right]}

\definecolor{lightgray}{cmyk}{0.3,0.3,0.3,0.3}
\definecolor{lightgray2}{cmyk}{0.1,0.1,0,0.1}

\newcommand{\np}{N_{\rm pix}}

\begin{document}

\hfill{ULB-TH/13-15}
\vskip 2cm
\centerline{
\Large{\textbf{ On the Scale of New Physics in Inflation}}}
\vspace{.6cm}
\centerline{\large{Lotfi Boubekeur}}
\vspace{0.1cm}
\begin{center}
\small{\sl
Abdus Salam International Centre for Theoretical Physics\\
Strada Costiera 11, 34151, Trieste, Italy.\\[0.2cm]
Service de Physique Th\'eorique, Universit\'e Libre de Bruxelles\\
Boulevard du Triomphe, CP225, 1050 Brussels, Belgium. \\[0.2cm]
Laboratoire de physique math\'ematique et de physique subatomique\\
Universit\'e de Constantine I, Constantine, Algeria.\\[0.5cm]
}
\end{center}
\vspace{1.5cm}
\noindent\hrule \vspace{0.3cm}
\noindent 
\small{\textbf{Abstract}}\\[3mm]
Effective field theory is a powerful organizing principle that allows to describe physics below a certain scale model-independently. Above that energy scale, identified with the cutoff, the EFT description breaks down and new physics is expected to appear, as confirmed in many familiar examples in quantum field theory. In this work, we examine the validity of effective field theory methods applied to inflation. We address the issue of whether {\it Planck-suppressed} non-renormalizable interactions are suppressed enough to be safely neglected when computing inflationary predictions. We focus on non-derivative non-renormalizable operators and estimate the cutoff that should suppress them using two independent approaches:  {\it (i)} the usual unitarity and perturbativity argument, {\it (ii)} by computing the UV-divergent part of the inflaton entropy, known to scale as the square of the UV-cutoff.  We find that in the absence of gravity (decoupling limit) the cutoff appears to depends linearly on the total inflaton excursion.  On the other hand, once gravity is restored, the cutoff is brought back to the Planck scale. These results suggest that inflationary scenarios with super-Planckian excursion are not natural from the EFT viewpoint. 
 
\vspace*{0.3cm}
\noindent\hrule \vspace{0.3cm}

\newpage

\tableofcontents

\section{Introduction}

Inflation is a UV-sensitive theory. Inasmuch this is considered as a  blessing, for it can probe much higher energy scales than any collider experiment, it can also be a curse, as predictions are sensitive to that same unknown energy scale(s). Such sensitivity manifests itself concretely in a variety of ways, the most common and severe one being the famous $\eta$-problem. Besides making inflation an unnatural theory, by requiring a non-trivial amount of fine-tuning, it also undermines both its remarkable predictive and observational successes.  

But we know that the presence of unknown high scale physics should not be a problem, as we can always parametrize our ignorance in the context of an effective field theory (EFT). The situation is 
however different in inflation mainly because gravity is non-renormalizable. In contrast with the usual case, focusing on the lowest dimension operators might not be enough, especially when the inflaton excursion exceeds the natural gravity cutoff scale; the Planck scale\footnote{Throughout the paper, we work in natural units $\hbar=c=1$. In these units, the reduced Planck scale is $M_P=1/\sqrt{8 \pi G_N}\simeq 2.44\times 10^{18}$ GeV, while the Planck mass is $m_P=G_N^{-1/2}\simeq 1.22\times 10^{19}\gev$. Our metric has signature $(-+++)$.} as we will argue below. 

In order to explain this point, let us borrow a useful analogy from particle physics.  In the standard model of strong and electroweak interactions, lepton number is a good quantum number respected by the renormalizable Lagrangian.  It is nevertheless broken at the non-renormalizable level through the dimension-five seesaw operator \cite{wop, seesaw} ${\cal O}_5=c_5 \,(\ell_L\cdot h)^2/M$, where $\ell_L$ is the lepton doublet, $h$ the standard model Higgs and $M$ a cutoff scale. Once the electroweak symmetry is broken, neutrinos pick up masses of ${\cal O}(v^2/M)$, where $v$ is the VEV of the SM Higgs. For $M\sim 10^{14}$ GeV, and $c_5\sim {\cal O}(1)$, neutrino masses are of the right order magnitude to explain observed neutrino oscillations. In order for this EFT to make sense, higher-order operators should be suppressed below the cutoff. Indeed, operators containing derivatives will be suppressed as powers of $E/M$, that is at energies below the cutoff $M$, it is safe to neglect them. Furthermore, because the Higgs VEV is so small with respect to the seesaw scale, higher-order operators \eg ${\cal O}_6=c_6 \,(\ell_L\cdot h)^2 (h\,h^\dagger)/M^3$ are further suppressed and thus only contribute negligibly to neutrino masses.  However, if now hypothetically we crank up the Higgs VEV to values $v\sim M$, then the next-to-leading operators like ${\cal O}_6$ would give significant corrections to the seesaw operator ${\cal O}_5$, and so on for  higher-order operators. This simply signals the breakdown of EFT  as the perturbative expansion cannot be trusted anymore. Note that this can happen even if derivative operators are adequately suppressed by the ratio $E/M$. In inflation, the natural cutoff is the reduced Planck scale. As cosmological predictions are computed when $E\sim H_\star$, where $H_\star$ is the Hubble rate at horizon crossing, operators containing space-time derivatives are expected to scale as powers of $H_*/M_P$, which is in fact smaller than one. On the other hand, non-derivative operators will scale as $\phi/M_P$, where $\phi$ is the inflaton. If this dimensionless ratio is large \ie $\phi\gg M_P$, which is in fact expected \cite{Boubekeur:2012xn} in the simplest models of inflation compatible with observational data \cite{Planck},  then this also signals the breakdown of EFT. Notice again that this happens despite the fact that $H_\star/M_P$ is small.

This {\it impasse} sounds like a fatal blow for inflation. The successful inflationary predictions, backed up by the impressive amount of data accumulated up to now, would be completely irrelevant as long as the underlying theory used to derive them is inconsistent. One can think of two possible way-outs to this problem. The first approach is just to ignore higher-order operators due to our lack of understanding of Planck scale physics. This is the approach usually assumed tacitly in most studies. However, besides going against the basic principles of EFT, it implicitly implies an infinite amount of fine-tuning; the fine tuning of all the {\it infinite} tower of non-renormalizable operators, which is far from being natural, not to mention feasible. The second approach consists in building models with a known UV completion (See \eg \cite{Burgess:2013sla} and reference therein.). In such a top-down approach, higher-order operators are calculable, but only for specific models. This is certainly more attractive than the first approach, however again, it does not conform to the crucial model-independent character of EFT. Of course, this approach is undeniably useful and can serve as an existence proof for specific constructions, however it lacks the power of EFT, which is well-documented in virtually all areas of physics.

In this paper, we will examine the validity of EFT methods used in studying inflation, by determining the cutoff scale by which higher order non-renormalizable operators are suppressed. In general, this scale is also expected to be associated with the appearance of new degrees of freedom, or new physics\footnote{In scenarios with small speed of sound, (weakly-coupled) new physics is expected to arise close to the horizon scale \cite{Baumann:2011su}.}, that come into play to make the description viable in the UV. In particular, we will focus on  higher order non-derivative operators of the type
\be 
{\cal L}^{(n+4)}_{\rm NR}={\phi^{n+4}\over \Lambda^n}\,,  \, {\textrm{with }} n\ge 1\, , 
\label{nr}
\ee
which are expected to be present in the Lagrangian on very general grounds. At this point it is important to emphasize that \eq{nr} is a {\em tree-level} term, and {\it not a perturbative quantum correction}\footnote{These corrections should not be confused with non-perturbative quantum gravitational effects, that are expected to break all global symmetries producing terms $c_n {\phi^{n+4}/ M_P^n}$,  where in general $c_n\sim e^{-S}$, with typical values of the action \cite{Kallosh:1995hi} $S\sim {\cal O}(10)$. It is worth noticing that, for inflationary scenarios with super-Planckian excursions, these corrections are not suppressed enough to be neglected. }, in contrast with what is argued in many studies (See \eg Chapter 2.4 of \cite{Linde:2005ht}). If shift symmetry were to be respected by the {\it whole} Lagrangian, then this type of contribution would be a quantum correction and hence will be suppressed as argued in \cite{smolin}. However, this is not the case here as these contributions are, from the EFT standpoint, allowed by all the symmetries of the Lagrangian and there is absolutely no seemingly consistent reason to discard them. In fact, neglecting \eq{nr} sharply   contradict the basic tenets of EFT. 

Now, let us give our motivation for considering this peculiar type of operators among all the available ones.   A different class of higher-order operators, has been the subject of many interesting studies (see \eg \cite{Creminelli:2003iq, Assassi:2013gxa,Avgoustidis:2012yc}), especially in connection with non-Gaussianity. These studies focused on  operators respecting the shift symmetry $\phi\mapsto\phi+c$, which plays an important r\^ole in inflation. Now, unless this symmetry is local, which is not the case in general, it is unprotected against quantum gravitational effects \cite{Kallosh:1995hi, Banks:2010zn}. Hence it is expected to be broken at the non-renormalizable level. Then, according to whether shift symmetry is a good symmetry of the renormalizable Lagrangian or not, there are two cases to be contemplated: 
\begin{itemize}
\item[-] {\em The shift symmetry is broken in the renormalizable Lagrangian.} Therefore there is no reason to expect it to be unbroken in the non-renormalizable Lagrangian. As a result, the contributions \eq{nr} will be present and will be as unsuppressed as their counterpart in the renormalizable Lagrangian. For instance, one can build models where the shift symmetry is broken spontaneously by the VEV of a spurion field coupled to the inflaton, in such a way that the Lagrangian is invariant under some global discrete symmetry, chosen to restrict the form of the inflationary potential. Writing all invariant terms, regardless of their dimension, and integrating-out the spurion through its VEV, will result in the breaking of the shift symmetry, occurring equally in both the renormalizable and non-renormalizable Lagrangian.

\item[-] {\it The shift symmetry is unbroken by the renormalizable Lagrangian.} This can be the consequence of an accidental symmetry, like for instance baryon and lepton number in the SM. However, again, unless shift symmetry is a gauge symmetry, it will be broken by quantum gravitational effects \cite{Kallosh:1995hi,Banks:2010zn} producing \eq{nr}, where we expect $\Lambda\sim M_P$. Then, one can always build renormalizable operators by contracting the legs of the non-renormalizable couplings\footnote{This observation is due to the late S. Coleman in \cite{Kamionkowski:1992mf}. See Fig.~1 for a schematic illustration. }.  Thus, the breaking propagates to the renormalizable Lagrangian. No matter how tiny is the non-renormalizable coupling, the cutoff suppression is compensated by powers of the cutoff coming from loop integration, leaving only loop factors. These last would eventually be compensated by the number of contributing non-renormalizable interactions.

\begin{figure*}[!t]
\begin{center}
\includegraphics[scale=2]{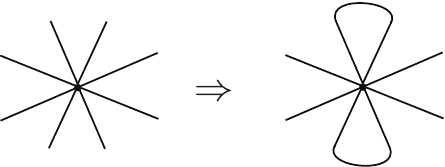} 
\end{center}
{\footnotesize{{\bf Figure 1: }A typical example of how the breaking of shift symmetry propagates from ${\cal L}^{n+4}_{\rm NR}$ to the renormalizable Lagrangian. By contracting the legs of the  $d=8$ non-renormalizable  $\phi^8/\Lambda^4$ operator, one obtains a renormalizable dimension-four effective operator breaking the shift symmetry in the renormalizable Lagrangian. Notice that the renormalizable Lagrangian receives contributions from {\it all} the non-renormalizable operators.}}
\end{figure*}

\end{itemize}
From the above discussion, it is clear that the set of operators \eq{nr} is expected to be present in the Lagrangian regardless of the fact that shift symmetry is broken at the renormalizable level or not. We have seen that owing to the non-renormalizable nature of the theory, and to the fact that the shift symmetry is global\footnote{This is valid as long the shift symmetry does not descend from a gauge symmetry that is conserved in the UV.}, one cannot seclude the breaking of the shift symmetry in either the renormalizable or non-renormalizable Lagrangian. This is unlike what happen happen in  a renormalizable theory, where quantum corrections will give divergences that have the same functional form of the starting Lagrangian. These divergences are regularized by including  only renormalizable counterterms in the Lagrangian. We have also seen that these operators will become important as soon as $\phi\gtrsim\Lambda_n$ and thus cannot be neglected evethough $H/M_P$ is small. Thus, in a consistent EFT description of inflation, one should include all shift symmetry breaking terms, both renormalizable and non-renormalizable on equal footing.  In the following, we will consider the shift symmetry as the underlying symmetry to formulate the EFT of inflationary dynamics.

The paper is organized as follows.  We will start with some heuristic arguments on the expected size of the cutoff in \S \ref{sec2}. In \S\ref{sec3}  we study to some depth the shift symmetry which we consider  later in as the underlying symmetry in formulating the EFT. Next, in \S \ref{sec4}, we compute the cutoff using unitarity and perturbativity. In \S \ref{sec:4}, we estimate the cutoff by computing the inflaton entropy. In \S \ref{sec6}, we comment on the connection with Wald entropy, and we derive a cutoff scale from its  consistency. We end up by discussing the results and laying down the conclusions in \S \ref{sec7}. Some useful details are collected in the appendices.

\section{Heuristics}
\label{sec2}
Effective Field Theory (EFT) is a powerful organizing principle that allows to study physical systems by focusing only on the low-energy degrees of freedom. By writing the lowest dimension operators compatible with the symmetries of the system, one can compute observables up to the desired accuracy and for energies below a certain cutoff, without committing to a specific model. Such cutoff scale can be determined requiring the EFT to be meaningful below that scale. A well-known way to do that is by demanding unitarity of elastic scattering amplitudes.  Unitarity is not only useful in determining the energy domain where the EFT makes sense, but it can also be used to constrain new physics; for instance deriving bounds on the mass of new degrees of freedom. Classical examples in particle physics include masses of the Higgs and top quark  \cite{Marciano:1989ns}. While in cosmology, a celebrated example is the upper bound on the WIMP mass \cite{Griest:1989wd}. 
In the same spirit, the EFT of inflationary fluctuations \cite{Cheung:2007st} makes the task of studying inflationary predictions in a model-independent way possible.  Like in particle physics, one can estimate the UV cutoff of EFT of inflationary fluctuations by considering the unitarity of scattering amplitudes. For instance, the $s$-wave unitarity of $2\to2$ scattering amplitude gives a non-trivial {\it theoretical bound} on the speed of sound \cite{Cheung:2007st} $c_s\gg 0.003$, which in hindsight is just the value of $c_s$ yielding an ${\cal O}(1)$ non-Gaussianity. The actual {\sl Planck 2013} data  \cite{Ade:2013ydc} give $c_s\ge 0.02$ at 95$\%$ CL, very close to the unitarity bound\footnote{See also \cite{Baumann:2011dt} for a discussion of the strong coupling in models with scale-invariant spectrum.}.

In the context of EFT of the inflationary background \cite{Weinberg:2008hq}, one can also study inflationary backgrounds model-independently. In \cite{Weinberg:2008hq}, Weinberg argue that one should consider non-renormalizable terms in the EFT of inflation even though derivative expansion appears to be trustworthy. Let us, for clarity, go through his line of arguments. First, at horizon exit, the fact that the ratio $H_\star/M_P\sim r^{1/2}$, where $r$ is the usual tensor-to-scalar ratio, is  small makes the use the renormalizable Lagrangian, with minimum number of space-time derivatives, to compute observables a very good approximation. However, a natural choice for the EFT cutoff is provided by the inflaton excursion during $\sim$ one e-fold at horizon 
crossing\footnote{From the definition of $\Lambda$, it is clear that it cannot exceed the Planck scale. In addition, using the definition of the power spectrum of curvature perturbation $\Delta_\zeta=H/(M_P 2\pi\sqrt{2\epsilon})$, this lower bound of this cutoff scale numerically corresponds to 
$$
\Lambda\gtrsim \sqrt{2\epsilon} M_P=\Delta^{-1}_\zeta \cdot H/2 \pi \sim 3.4\times 10^3 H\, , 
$$
which is of the same order of the bound derived in \cite{Assassi:2013gxa} from non-Gaussianity from dimension-five mixing operators for $\fnl\sim {\cal O}(1)$.} 
$\Lambda\gtrsim{\dot\phi_c\cdot H^{-1}}=\sqrt{2\epsilon} M_P $. It follows that derivative expansion is, in fact, under control since the expansion parameter always satisfies $H/\sqrt{2\epsilon} M_P\lesssim 10^{-5}$. In contrast, non-derivative non-renormalizable terms  \eq{nr}, which should be suppressed {\it at least} by this scale, \ie $
{\phi^{n+4}/ \Lambda^n}\lesssim {\phi^{n+4}/ \({\sqrt{2\epsilon} M_P}\)^n}$, and thus can be significantly unsuppressed with respect to their derivative counterparts, especially for large inflaton displacements. To see this, one can impose two theoretical consistency constraints. The first one is the validity of semi-classical methods \ie  ${\cal L}^{(n+4)}_{\rm NR}\ll M_P^4$ which gives 
\be
\Delta\phi\lesssim\phi\ll M_P \cdot \epsilon^{n\over 2(n+4)}
\label{eq1}
\ee

The second constraint is simply perturbativity: non-renormalizable terms should not overwhelm the tree-level potential for energies beneath the cutoff. 
\be
\Delta\phi\lesssim\phi\ll M_P \(\frac{H}{M}\)^{2\over n+4}\cdot \epsilon^{n\over 2(n+4)}
\label{eq2}
\ee
Both constraints \eq{eq1} and \eq{eq2} seem to suggest that the validity of EFT implies that excursions should be sub-Planckian.  This is also in  agreement with the unitarity requirement obtained from the interaction \eq{nr}
\be
\Delta\phi\le\({\pi\over 3}\)^{1/n}\, \Lambda\simeq \Lambda\, , 
\label{un}
\ee
where, once again, we are barring accidental cancellations with  other contributions. If we make the natural identification $\Lambda\simeq M_P$, we reach the same conclusion \ie  a perturbative and unitary  description of inflation requires that $\Delta\phi\lesssim M_P$, which  contradicts the observation made \eg in \cite{Boubekeur:2012xn} that most successful single-field models, accomplishing enough e-folds and matching observation entails the inflaton to roll over super-Planckian distances \ie $\Delta\phi\gg  M_P$. 

We have learned from this simple heuristic estimates that: \begin{itemize}
\item[{\it (i)}] The inflaton excursion and the EFT cutoff are directly linked \ie $\Delta\phi\lesssim \Lambda$. Combined with the fact that  the Planck scale is the natural cutoff of the EFT in inflationary theories \ie 
\be
\Delta\phi\lesssim M_P\, .
\ee

\item[{\it (ii)}] On the other hand, Typical single-field scenarios, including chaotic and hilltop inflation \cite{hilltop,Boubekeur:2012xn}, have 
\be 
\Delta\phi\gg M_P\, , 
\ee
leading to a direct contradiction with  {\it(i)}.

\end{itemize}

In the remainder of the paper, we will make the case for this statement more precise by computing the corresponding UV cutoff and comparing it with the Planck scale.

\section{Interlude --- The shift symmetry: the inside story}
\label{sec3}
Successful inflation demands that the shift symmetry $\phi\mapsto\phi+\Delta\phi$, where $\Delta\phi$ is a {\em constant} with dimension of mass, holds as  an approximate symmetry during inflation. Such a requirement is just expressing that the potential is (almost) flat between inflaton values $\phi_i$ and $\phi_f$. Let us for the moment assume that the shift symmetry is valid for all inflaton values \ie $\Delta\phi$ arbitrary. The corresponding Noether current is thus\footnote{The prefactor $i$ is immaterial, however as we will see below, it is essential in order to make the VEV $\vev{0|\[Q,\phi\]|0}$ real, as it should be.  See \eq{vevp}. Also, notice that $\Delta\phi$ is included explicitly in the definition of ${\cal J}^\mu$ because this last should have dimension length$^{-3}$.}
\be
{\cal J}^\mu=i {\delta {\cal L} \over \delta {\del_\mu \phi}}\cdot \Delta\phi\,, 
\ee
where ${\cal L}$ is the Lagrangian of the canonically normalized inflaton\footnote{We will focus on canonically-normalized single-field inflation scenarios as they are favored by current bounds on non-Gaussianity and isocurvature modes. }. In the absence of a non-trivial potential, ${\cal J}^\mu$ is trivially conserved as a consequence of the inflaton equation of motion. In this case, the  Noether charge 
\bea
Q=\int \d^3 x \, {\cal J}^0=i \int \d^3 x \, \dot\phi\, \Delta\phi\,  \eea
is conserved in time \ie $\dot{Q}=0$. On the other hand, the shift symmetry is broken by the presence a potential with non-vanishing tilt, and the charge is no more conserved as $\dot{Q}\propto \Delta\phi\, V'$. In addition to the tilt of the potential, the breaking of the shift symmetry is also weighted by $\Delta\phi$, which now takes a finite value. 
Using the canonical commutation relations
\be
\[\dot\phi(\vec{x}, t),\,\phi(\vec{y}, t)\]=-i
\delta^{(3)}(\vec{x}-\vec{y})\,, 
\label{comm}
\ee
one gets the variation of $\phi$ under shift symmetry
\be
\delta_Q\phi=\[Q(t),\,\phi(t)\]=\Delta\phi(t)\,, 
\label{eq11}
\ee
where $\Delta\phi(t)$ is just the shift in the inflaton at time $t$, which can also be written as 
\be 
\Delta\phi(t)\equiv\phi(t)-\phi(0)= \int_0^{t} \d t'\,  \del_{t'}{\phi}_c(t')
\,, 
\ee
where we have taken $t=0$ as a reference initial time and $\phi_c$ is the inflaton homogeneous background value. For exact shift symmetry, the time-derivative of \eq{eq11} vanishes \ie $\dot\phi_c=0$, where we kept the leading order in slow-roll parameters.  On the other hand, this is no more the case once shift symmetry is broken. As in usual gauge theory \cite{Goldstone:1962es}, the order parameter signaling the breaking is just the VEV of the Hermitian operator \eq{eq11}
\be 
\vev{0|\[Q(t),\phi(t)\]|0}=\Delta\phi(t)\,, 
\label{vevp}
\ee
which in other words means the vacuum is no more annihilated under the action of the charge\footnote{In flat spacetime, it is easy to see that  a new state, labeled with $\Delta\phi$, can be obtained by acting upon the vacuum as follows 
$
|\Delta\phi\rangle=e^{i Q}|0\rangle=e^{{-i\Delta\phi/ 2}( a_0- a^\dagger_0)}|0\rangle$, where $a_0\equiv a_{\vec{k}=0}$ is the zero-mode annihilation operator.  Such a state is different from original vacuum since it satisfies $\vev{0|\Delta\phi}=0$, and is in fact a zero-momentum coherent-state \ie a Nambu-Goldstone boson.} $Q$.

From the canonical commutation relations \eq{comm}, one can conclude that for non-vanishing $\dot{\phi_c}$, the charge $Q\sim r$ as $r\to \infty$ and so it does not fall-off rapidly enough as it should for a well-defined Noether charge. This is just another way to see that the shift symmetry is spontaneously broken for non-vanishing inflaton velocity. Using the Goldstone theorem, we can determine the breaking scale\footnote{Alternatively, we can get the same result repeating the argument of Sec.(2.4.1) of \cite{Baumann:2011su} considering contact terms $\vev{0|{\rm T}\{e^{ikx}{\cal J}^\mu (x){\cal J}^\nu(0)\}|0}$. 
}
 $f_\pi$ of the shift symmetry by evaluating the matrix element\footnote{In what follows and for consistency with current algebra literature, we switch to the definition of the current ${\cal J}^\mu$ without the prefactor $i$. }
\be
\vev{0|{\cal J}^\mu(x)|\pi(k)}=-i f_\pi\, k^\mu\,e^{-ikx}\, , 
\label{goldstone}
\ee
where as usual, the $p^\mu$ factor is expected from Lorentz invariance and non-vanishing of the above matrix element signals the breaking of shift symmetry. It also signals the existence of a state $\pi$ that can be created from the vacuum through the action of the current ${\cal J}^\mu$; the Nambu-Goldstone boson. Expanding around the homogeneous shift symmetry breaking solution $\phi(t)=\phi_c(t)+\pi(x)$, and using the normalization $\vev{0|\pi(x)|\pi(p)}=e^{-ip.x}$, one gets 
\be
f_\pi=\Delta\phi\,.
\ee
\begin{figure*}[!t]
%\begin{center}
\vspace*{-10mm}
%\begin{tabular}{cc}
\hspace{-5mm}
\includegraphics[scale=1.25]{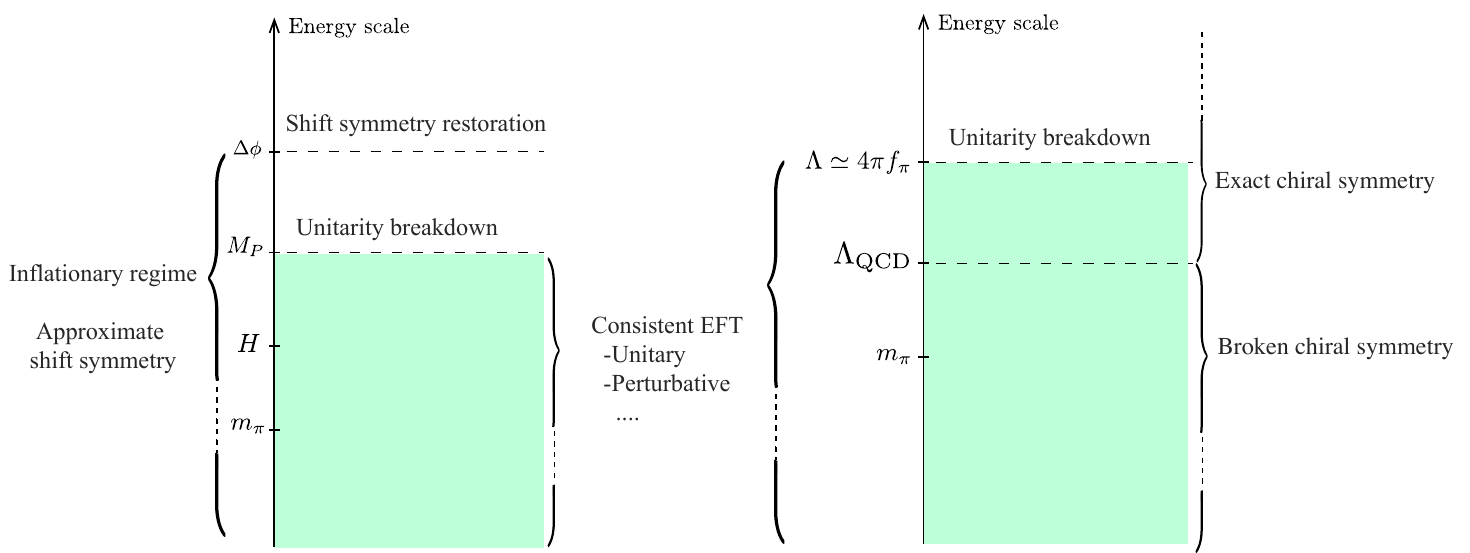} 
%\end{tabular}
%\end{center}
{\footnotesize{{\bf Figure 2: }The energy scales involved in the EFT (not to scale). In the right panel, we represent the familiar case of the EFT of pions. In the left panel,  we represent inflationary EFT, for the case $\Delta\phi\gg M_P$, \ie scenarios with super-Planckian excursions.  Beneath $\Delta\phi$, the shift symmetry is spontaneously broken, and is realized as an approximate symmetry during inflation. Above $\Delta\phi$, the shift symmetry is restored. In both cases, the green area stands for healthy EFT.}}
\end{figure*}
As in gauge theory, broken symmetries get restored at energies above the breaking scale. In terms of the matrix elements \eq{goldstone}, this means that there will be additional terms,  powers of $(E/f_\pi)$, contributing to the rhs and eventually canceling it.\footnote{A  textbook example to illustrate this consists of a complex scalar field whose Lagrangian \be{\cal L}=-\frac12|\del\phi|^2 - \frac\lambda2(|\phi|^2-v^2)^2\ , \ee is invariant under a global $U(1)$ transformation $\phi\mapsto e^{i\theta}\phi$, where $\theta$ is a real number. Spontaneous symmetry breaking produces a massive particle with mass $m_\rho=\lambda^{1/2} v$ and a massless Goldstone mode $\pi$. The corresponding Noether current reads $J^\mu= f_\pi \cdot \del^\mu\pi$ with $f_\pi\equiv v$. For energies below $m_\rho$, we can integrate the $\rho-$particle, and obtain an effective Lagrangian for the $\pi$'s 
$$
{\cal L_\pi}=-\frac12(\del\pi)^2-\frac{\lambda}{2 m_\rho^4}(\del\pi)^4+\cdots$$As long as $E<v$, the additional terms are negligible and do not contribute significantly to the current. However, as energy approaches $v$, {\it all} the additional terms dominate with the same magnitude. This is just the manifestation that the original $U(1)$ is restored, and the that above effective Lagrangian is no more valid.}
Likewise, for the case at hand, we expect the same to happen \ie for energies above $\Delta\phi$, shift symmetry will be restored. While at energies below $\Delta\phi$, shift symmetry is broken and the expansion parameter of the EFT is\footnote{Here we are considering the standard dispersion relation $E\sim k$. However in general, the expansion parameter is $k/\Delta\phi$. Moreover, when computing cosmological observables, as usual $k$ is understood as the wavenumber where perturbations freeze, \ie at horizon crossing, that we denote as $k_\star$ from now on.}  $E/\Delta\phi$. In particular, for cosmological correlation functions, the expansion parameters is $H/\Delta\phi$ and is consistently smaller than one. As an example, consider the two-point function of the primordial curvature perturbation $\zeta$, which can be estimated as
\be
\vev{\zeta^2}\sim\[\frac{k_\star}{\Delta\phi(t=H^{-1})}\]^2\, . 
\label{eq:zeta}
\ee
In the case of standard slow-roll inflation, $k_\star\simeq H$ and the denominator of \eq{eq:zeta} equals to $(\dot\phi_c\cdot H^{-1})^2$, reproducing the well-known textbook result  $\vev{\zeta^2}^{1/2}\sim H/M_P\sqrt\epsilon$. This estimate also applies in the presence of non-standard dispersion relations. For instance,  in ghost inflation \cite{ArkaniHamed:2003uz}, $E\sim k^2/M$ and $k\sim (M^3 H)^{1/4}$, where $M\equiv\dot\phi_c^{1/2}$ is the mass scale suppressing the non-standard kinetic term. From this one gets the correct scaling $\vev{\zeta^2}^{1/2}\sim (H/M)^{9/4}$.

One can also compute the mass of $\pi$; the Pseudo Nambu-Goldstone Boson (PNGB), by using 
\be
\vev{0|\nabla_\mu{\cal J}^\mu(0)|\pi(k)}=- f_\pi m_\pi^2\, . 
\ee
Using the inflaton equation of motion $\ddot{\phi}+3H\dot\phi=-V'$, we recover the expected result\footnote{This is in agreement with the general formula of the PNGB mass matrix (See \eg Eq. (19.3.20) of \cite{Weinberg:1996kr}.)
$$
m^2_{ab}=-f^{-1}_{ac} f^{-1}_{bd}\[Q_c,\[Q_d,H_1\]\] \,,
$$ 
where $H_1$ is the Hamiltonian breaking explicitly the symmetry and the $f_{ab}$ are defined as a generalization of the Goldstone theorem \eq{goldstone} to the case with several PNGBs $$\vev{0|{\cal J}_a^\mu(x)|\pi_b (k)}=-i f_{ab}\, k^\mu\,e^{-ikx}\,, $$ 
corresponding to the breaking of several charges $Q_a$. In our case, $H_1\equiv V$ and there is a single PNGB, with "decay constant" $f_\pi$.}
\be
m_\pi^2= V''(\phi_c)\,
\label{mpi}
\ee
The EFT defined above shares many features with the familiar example of QCD pions (See Fig.~2).  As any consistent EFT, it is characterized by two scales. The first one is the breaking scale of the symmetry, which in our case is the breaking scale of the shift symmetry $\Delta\phi$, describing the scale of the flatness of the potential. The explicit breaking of the shift symmetry is accompanied, as usual, with the appearance of a PNGB. The mass of this last is given by \eq{mpi}, naturally below $H$, by virtue of the slow-roll conditions.  Therefore, it will be captured in the EFT because the cutoff should be, in any case,  larger than $H$. The second important scale is the cutoff; the energy scale beyond which the EFT breaks down, which we will estimate in the next section.

As usual, one can reintroduce the Goldstone boson through the Stueckelberg trick, where now $\phi$ and $\pi$ obey the transformation rule
\be
\phi\mapsto\phi+\xi \quad \textrm{and~~~}\pi\mapsto\pi-\xi\ , 
\ee
one can write the renormalizable Lagrangian 
\be
{\cal L(\phi+\pi)}=-\frac12 g^{\mu\nu}\del_\mu(\phi+\pi)\del_\nu(\phi+\pi)-V(\phi+\pi)\,, 
\label{eqq:20}
\ee
which is invariant under the above transformations. Decomposing the metric using the ADM formalism as $\d s^2= -N^2 \d t^2+h_{ij}(\d x^i + N^i \d t) (\d x^j + N^j \d t)$ where\footnote{This is the so-called $\phi$-gauge, as opposed to the well-known $\zeta$-gauge \cite{Maldacena:2002vr}.} $h_{ij}=a^2(t)(\delta_{ij}+{\gamma_{ij}\over M_P}+\cdots)$ and expanding the Lagrangian, we get the mixing
\be
{\cal L}_{\rm mix}=-h^{ij} N_i\dot\phi\,\del_j\pi
\label{lmix}
\ee
where $N^i$ is determined by solving the constraints equations \cite{Maldacena:2002vr} yielding $N^i=\del_i \chi$ with 
\be 
\chi=-\del^{-2}\[{\dot\phi^2\over 2 H^2}{\d\over\d t}\(\frac{H}{\dot\phi} \pi\)\]. 
\ee
From \eq{lmix}, it is easy to see that the tensor-scalar mixing is irrelevant as long as the slow-roll conditions $\epsilon\ll 1$ is satisfied. This leads to the well-known observation that at quadratic order, tensor and scalar perturbations evolve independently. Furthermore, the quadratic action of the PNGB $\pi$ is just the action of a free scalar field with mass\footnote{Expanding the Lagrangian \eq{eqq:20} to quadratic order, one gets
$$
S=\int \d^4 x \sqrt{-g}\(-\frac12({\del \pi})^2-\frac12({\del \phi})^2 + \dot\phi\,\dot\pi -V(\phi)-\pi V'(\phi)-{\pi^2\over 2} V''(\phi) \)\,.
$$
Now, integrating by parts the mixing term $\dot\pi\dot\phi$ and using the classical equation of motion for $\phi$, the tadpole term $\pi V'(\phi)$ cancels exactly, leaving only the mass term $\pi^2 V''(\phi)/2$.}
 \eq{mpi}. 
\section{The physical cutoff from unitarity and perturbativity}
\label{sec4}
In this section, we will determine the possible values of the UV cutoff $\Lambda$ using unitarity and perturbativity as a guiding principle.  Depending on the prescription and on the process/operator under consideration, the cutoff can take different possible values. Among these lasts, it is fair to conservatively identify the smallest as the physical cutoff, that is $\Lambda=\min\{\Lambda_i\}$. Moreover, sometimes the unitarity fails to give sensible constraints because either the scattering amplitude diverges (in the infrared), or is vanishing, so one has to rely on perturbativity to derive the UV cutoff.   
In both prescriptions, the energy scale $\Lambda$ is as usual associated with the appearance of new degrees of freedom, or strong coupling, or both. These new degrees of freedom are the ones that will restore unitarity and perturbativity.

Let us begin with perturbative unitarity. This is the most well-defined and robust prescription. The unitarity constraint requires that scattering amplitudes of some allowed scattering processes should be well-behaved at high energy. The maximum scale where this is satisfied is identified with the cutoff $\Lambda$. 
For the case at hand, expanding \eq{nr} around the classical background $\phi_c$, one gets the following quartic interaction
\be
{\cal L}^{(n+4)}_{\rm NR} \supset\frac{(4+n)!}{4!\,n!} \, \({\phi_c\over  \Lambda}\)^n \, \pi^4\, .
\ee
Unitarity of $2\to2$ scattering from the above interaction implies\footnote{One gets the same results in the so-called $\zeta$-gauge, through the transformation \cite{Maldacena:2002vr} $
\zeta=-\frac{H\pi}{\dot\phi}$ and considering the $2\to 2$ scattering of $\zeta$.} 
\be
\phi_c\le \Lambda \cdot \[{8\pi\, n!\over (4+n)!}\]^{1/n}
\label{eq:20}
\ee
Next, we consider the perturbativity constraint; where the cutoff is the scale at which perturbation theory breaks down.  In renormalizable theories, this cutoff scale is also related to the scale where the perturbative expansion of amplitudes cannot be trusted anymore. At this scale, the theory becomes strongly coupled and one-loop diagrams become as large as the tree-level ones. 
Expanding \eq{nr} around $\phi_c$ and keeping the trilinear term, maintaining perturbativity implies that the ratio\footnote{In the classic example of pions \be  {\cal L}=-\del_\mu\pi^a \del^\mu\pi^a +{1\over 6 f_\pi^2} (\pi^a\pi^a\del_\mu\pi^b\del^\mu\pi^b-\pi^a\pi^b\del_\mu\pi^a\del^\mu\pi^b)  +\cdots\,, \ee the tree-level perturbativity ${\cal L}_3<{\cal L}_2$ is satisfied for energies $E<f_\pi$. This is troublesome because $f_\pi$ is {\it lower} than $m_\pi$, meaning that pions would decouple and thus are not captured by the EFT. It turns out that this conclusion is wrong because the tree-level and one-loop amplitude are comparable, which results in a higher cutoff scale $\Lambda_\pi= 4\pi f_\pi$, rather than the naive $f_\pi$. Something similar happens in Higgs inflation, where looking at the $s$-channel contribution of $h h\to h h$ scattering, one is tempted to conclude that the cutoff is $\Lambda\sim M_P/\xi$, where $\xi$ is the non-minimal coupling. However including the $t$ and $u$-channels, the cutoff is brought back to $M_P\gg M_P/\xi$. See \eg discussion in \cite{Hertzberg:2010dc}. }
\be
{{\cal L}_3\over {\cal L}_2}{\Bigg{|}}_{E=H}
\simeq \(\phi_c\over \Lambda\)^{n+1}\cdot\(\Lambda\over H\)<1\, .
\label{ineq}
\ee
To capture the dynamics of the PNGB, one must have $\Lambda>H>m_\pi$. Therefore, the inequality \eq{ineq} translates into the following bound on the inflaton excursion 
\be
\Delta\phi\le\phi_c<\Lambda
\ee
From the above discussion, it is clear that there could be many different energy scales that could play the r\^ole of $\Lambda$. Although in general, these cutoffs will not coincide, they will be proportional.  For instance if we consider unitarity, the difference between the corresponding cutoffs is simply due to the partial wave contributing to the each  scatterings (See Appendix \ref{app:unit} for details). In the case of perturbativity, the difference is due to loop factors among other things.    

Let us now apply these prescriptions to determine the cutoff for the  inflationary Lagrangian. Applying the first prescription (unitarity), for the graviton mediated elastic scattering $\pi\pi\to\pi\pi$, one obtains as expected that the naive cutoff is $M_P$. We refer to Appendix \ref{app:unit} for details. This makes sense perfect as gravity will become strong at that energy scale.  Next, let us determine the cutoff using the second prescription (perturbativity). For the moment, let us ignore gravity by taking the {\it decoupling limit} $M_P\to \infty$ and ${\dot H}\to 0$ keeping both  $M_P^2{\dot H}$ and $H$ fixed.  Now, let us estimate the cutoff, which by definition is the {\it smallest} energy scale at which non-renormalizable operators become relevant. We can write the non-renormalizable potential \eq{nr} as\footnote{Notice that this factorization is not unique, but this is not very important for what follows.}
\be 
{\cal L}^{(n+4)}_{\rm NR}(\phi)=\widetilde{V}(\phi)\times \sum_{n=1}^{\infty} (\phi/\Lambda_n)^{n} \, , 
\label{sum}
\ee
where $\widetilde{V}(\phi)$ contain only dimension-four terms that are of ${\cal O}(M_P^2 H^2)$.   According to the standard d'Alembert test,  the sum in \eq{sum} will converge to a number of ${\cal O}(1)$ if  $\phi<\Lambda_n$.  It follows that the cutoff is $\Lambda\sim \Lambda_n\sim \Delta\phi$, where $\Delta\phi$ is the total excursion of the inflaton during the slow-roll phase. Notice that in above limit, it is safe to ignore quantum gravitational corrections to the potential. Therefore, at least in the above decoupling limit, it appears that the value $\Lambda\sim \Delta\phi$ is an equally consistent value for the cutoff. But even in this limit,  how can one understand such a conclusion? One way to interpret this result is the following. During inflation, the inflationary trajectory is a flat direction, which effectively has a large VEV of order $\Delta\phi$. While the inflaton  remains light $m_\phi\ll H$, whatever degree of freedom coupling to it will pick a huge mass, proportional to the inflaton VEV; so that\footnote{Here, we are restricting ourselves to the situation where the coupling to the other degrees of freedom gives rise to a {\it positive} mass contribution. We do not consider the more complicated case where the coupling makes heavy degrees of freedom lighter \ie $m_{\rm others}<H$. This also signals the breakdown of the EFT as these degrees of freedom should be taken into account in the EFT, in addition to the inflaton.} $m_{\rm others}\propto \Delta\phi$. At energies below the cutoff, one can integrate-out these heavy modes from the Lagrangian. This, as usual,  will in generate an infinite tower of non-renormalizable operators suppressed by the heavy mass scale 
\be
{\cal L}^{(n+4)}_{\rm NR}(\phi)=\sum_{n=1}^\infty \lambda_n \, \phi^4 \(\frac{\phi}{\Delta\phi}\)^n\, ,
\ee
where the Wilson coefficients $\lambda_n$ can be computed from the original Lagrangian. For ${\cal O}(1)$ coefficients the cutoff can be identified as $\Lambda_n\equiv\Lambda\simeq\Delta\phi$. This conclusion can also be reached using the unitarity argument. Upon integrating-out the heavy fields at one loop, one gets  that ${\cal A(\pi\pi\to\pi\pi)}\sim \frac{1}{16\pi^2}(E/\Delta\phi)^4$, where we ignored ${\cal O}(1)$ coefficients. The resulting unitarity cutoff is again $\Lambda\simeq\Delta\phi$.
\section{The physical cutoff from de Sitter entropy}
\label{sec:4}
Entropy is a very useful quantity to characterize physical systems. In particular, it encodes the amount of ignorance or lack of information in describing any given system. It is also related to the number of  degrees of freedom contained in the system.  As such, entropy is a UV-sensitive quantity; it grows with the number of degrees of freedom below some given cutoff. Black holes are the classic example of this statement, they possess an entropy given by the well-known Bekenstein-Hawking formula $S_{BH}=A/4G_N$, where $A$ is the horizon area. The Bekenstein-Hawking entropy scales as the square of the cutoff $M_P$, and represent the number degrees of freedom carried by the black hole on its horizon. We will take advantage of this property to estimate the cutoff for the EFT of inflation \cite{Conlon:2012tz}. 

The horizon of de Sitter (dS) space shares many properties with their black hole analogues. As for black holes, the dS metric in its static form has the remarkable property that it possesses a timelike Killing vector. Furthermore, the dS horizon is a bifurcate Killing horizon, \ie a surface where the norm of a Killing vector vanishes.  The similarity with black holes is more manifest if we consider dropping objects in dS.  For instance, if a massive object, held at a distance $R$ from a dS observer, is released, it will be attracted towards the horizon with an acceleration of $H^2 R$, until disappearing beyond it. Hence, it follows that if one waits enough time, all matter (degrees of freedom) will disappear beyond the horizon, leading to an information paradox similar to the black holes one. Like in the case of black holes, the resolution of this paradox lies is in the fact that the dS horizon carries an entropy which is equal to one quarter of the horizon area $A$ in units of the Planck length $\ell_P=G_N^{1/2}$. In the case of de Sitter, entropy is given by  \cite{Gibbons:1977mu} 
\be
S_{dS}={1\over 4} {A\over \ell^2_P}= 8 \pi^2 {M_P^2\over H^2}\,, 
\label{dsentropy}
\ee
similar to the Bekenstein-Hawking area law for black holes. The second law of thermodynamics states that the entropy of a closed system never decreases with time. The expanding universe during slow-roll inflation does not escape this rule.  For instance, in single-field inflation, the second law implies an absolute bound on the amount of tensor modes \cite{Boubekeur:2012xn}.

Also, given that entropy is related to the number of micro-states and therefore to the dimension of Hilbert space, expression \eq{dsentropy} suggests the tantalizing possibility that the Hilbert space of de Sitter is finite\footnote{ The authors of \cite{Goheer:2002vf} argue that finite entropy does not necessarily imply finite Hilbert space, but implies at most a discrete Hilbert space.}  (See discussion in \eg \cite{Witten:2001kn} and references therein). Conversely, one can think of \eq{dsentropy} as a definition of the cutoff of the theory in de Sitter with a given Hilbert space dimensionality. In this section, following \cite{Conlon:2012tz}, we shall discuss the entropy  contribution of the inflaton, and using the above connection we will derive the cutoff for a theory with scalar fields in de Sitter.

\subsection{Coarse grained entropy}
\label{sec:1}
\begin{figure*}[!t]
\begin{center}
\includegraphics[scale=0.6]{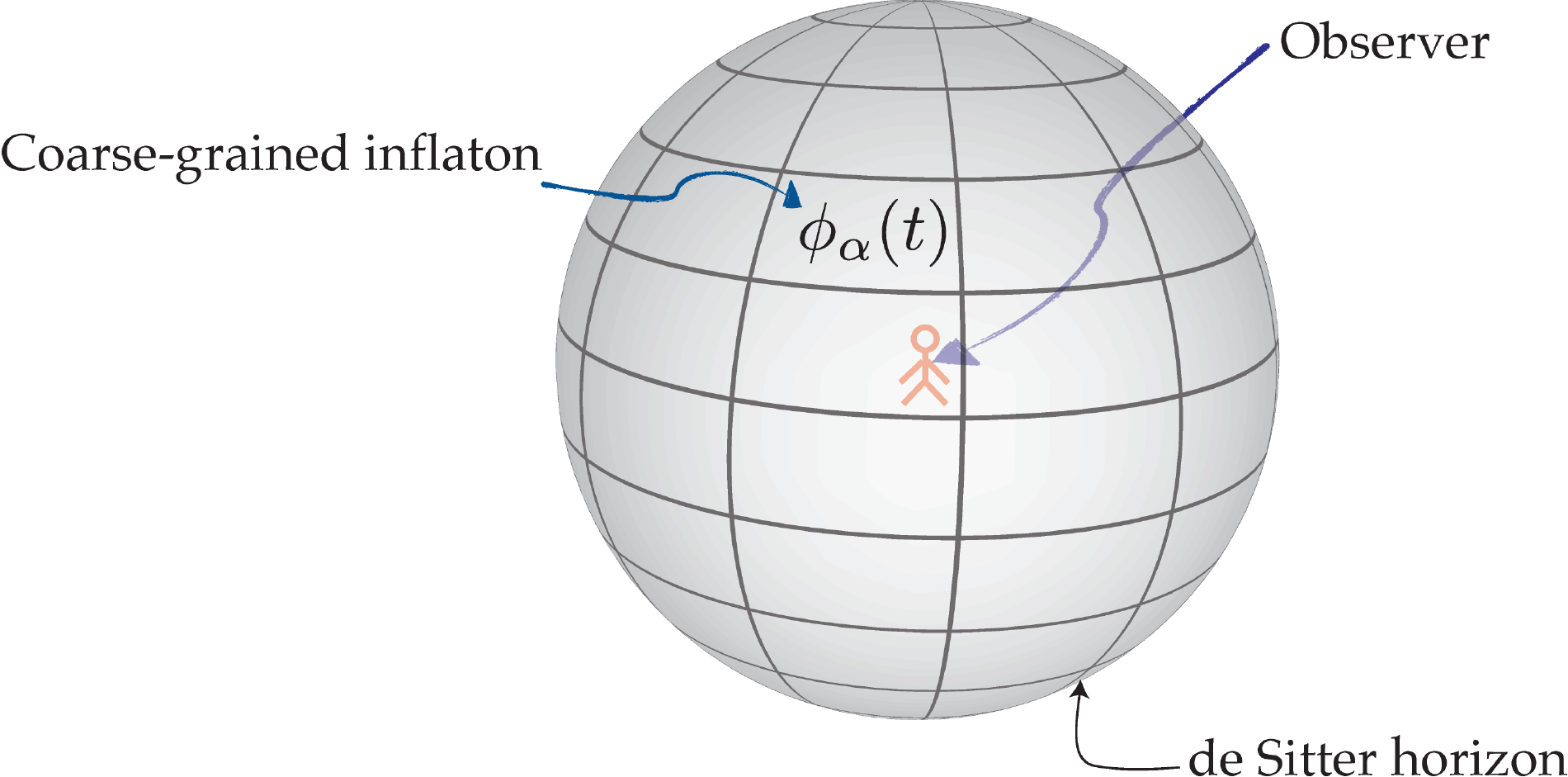} 
\end{center}
{\footnotesize{{\bf Figure 3: }Schematic view of the de Sitter horizon of an observer. Each pixel $\alpha$ has area $\ell^2_P$ and is assigned a coarse-grained inflaton $\phi_\alpha(t)$.}}
\end{figure*}
In the following, we will compute the inflaton entropy by counting the  the (micro-)states $\Omega$ that can live inside the horizon volume. According to horizon complementarity \cite{Goheer:2002vf}, it is enough to count states that live on the boundary. One way to do so is to calculate the inflaton statistical entropy associated with a given field configuration observed in our causal patch. For the case at hand, we can calculate $\Omega$ by counting the number of microscopic states corresponding to an asymptotic dS observed state. This state corresponds to a certain macroscopic configuration of the inflaton field $\phi(t_f, \,\vec{x})$, at time $t_f$, then $\Omega$ will correspond to how many microscopic configurations $\phi(t_i, \,\vec{x})$, with $t_i<t_f$ that can yield this observed final state.  In order to compute $\Omega$, we must therefore coarse-grain the horizon area and compute the corresponding entropy (See Fig.~3). 
We can discretize the horizon area into a lattice of a minimal spatial spacing $a_{\rm min}$, to be determined later. Then, the number of pixels $\np$ contained in the lattice will be  
\be
\np=\frac{A_{dS}}{a_{\rm min}^2}={4 \pi \over H^2 a^2_{\rm min}}\,.
\ee
In each pixel $\alpha$, there is a 2D scalar field $\phi_{\alpha}(t)$. For instance, for a square lattice, $\phi_\alpha(t)\equiv\phi_{i,j}(t)=\phi(t, i \,a_{\rm min}, j a_{\rm min})$, where $i,\, j=1, 2, \cdots, \sqrt{\np}$, which is just the discretized version of the continuum field.   In other words, the continuum field is just a long wavelength approximation of the discretized version.  The observed configuration $\phi(t_f, \vec{x})$ corresponds to and arrangement of $\np$ pixels that produces the same macroscopic configuration (scalar field profile).  Furthermore, to a certain configuration, there correspond many ways of tessellating the horizon with the $\phi_{i, j}$s. Conversely, the one observed represents only one possibility among the available ones.  On the other hand, it is easy to see that by swapping the $\phi_{i,j}$ does not change the observed macroscopic configuration. Notice that each $\phi_{i,j}$, on each one of the pixels, will evolve according to its classical equation of motion, which in the slow-roll approximation has the solution\footnote{We are assuming that $H\simeq$ constant, and so will be $\epsilon$. We additionally assume that $\epsilon$ is large enough to neglect quantum inflaton's fluctuations. See appendix \ref{cg} for details. } $\phi_{i,j}(t_f)=\phi_{i,j}(t_i)\pm \sqrt{2\epsilon} M_P H (t_f-t_i)$, where the $\pm$ sign corresponds to the opposite sign of the first derivative of the inflaton potential. 

Now, it is easy to compute the number of initial coarse-grained scalar field configurations. This corresponds to the number of independent initial conditions of the classical (slow-roll) equation of motions. Furthermore, we can also include the scalar fields quantum fluctuations that randomly kick these lasts up and down the inflationary potential. After $N$ e-folds, the net result is that at each e-fold, quantum fluctuations double the number of corresponding initial conditions. We refer to Appendix \ref{cg} for more details. The resulting number of  micro-states is then   
\be 
\Omega=2^N \, \np!
\label{omega}
\ee 
The $\np!$ factor is understood as the result of $\np!$ set of initial conditions that lead to the dS asymptotic state, while the $2^N$ factor is understood as the uncertainty introduced when considering the stochastic dynamics of the inflaton.  

Plugging our result \eq{omega}, in the Boltzmann entropy formula $S=\log\Omega$, one gets 
\be
S\simeq 
\np\log \np-\np + N \log2
\simeq 8 \pi \(1\over H^2 a_{\rm min}^2\) \log\(2 \pi^{1/2}\over H a_{\rm min}\)- \(\frac{4 \pi}{H^2 a^2_{\rm{min}}}- N\log2\)\,, 
\label{coarse}
\ee 
where we used the Stirling formula. Recall that in this setup, in order to make sense, the number of e-folds is defined as $N=\log(k_{\rm max}/k_{\rm min})$. Hence, since $k_{\rm max}=\Lambda$ and $k_{\rm min}=H$, then $N=\log(\Lambda/H)$. 

We can also express \eq{coarse} entirely in terms of the UV cutoff $\Lambda$ by writing the lattice spacing $a_{\rm min}$ in terms of $\Lambda$. On dimensional grounds $\Lambda$ should scale as $1/a_{\rm min}^2$. The precise proportionality constant is determined in Appendix \ref{app:amin}. Plugging \eq{cutoff} into \eq{coarse}, we get
\be
S \simeq 8 \pi^2 \[4\pi\(\Delta\phi\over H\)^2\cdot\; \log\(4 \pi^{3/2}\Delta\phi\over H\)- \(\frac{2 \pi\Delta\phi^2 }{H^2 }- {N\over 8 \pi^2}\log2\)\]\,. 
\ee 
\begin{figure*}[!t]
\begin{center}
\includegraphics[scale=0.6]{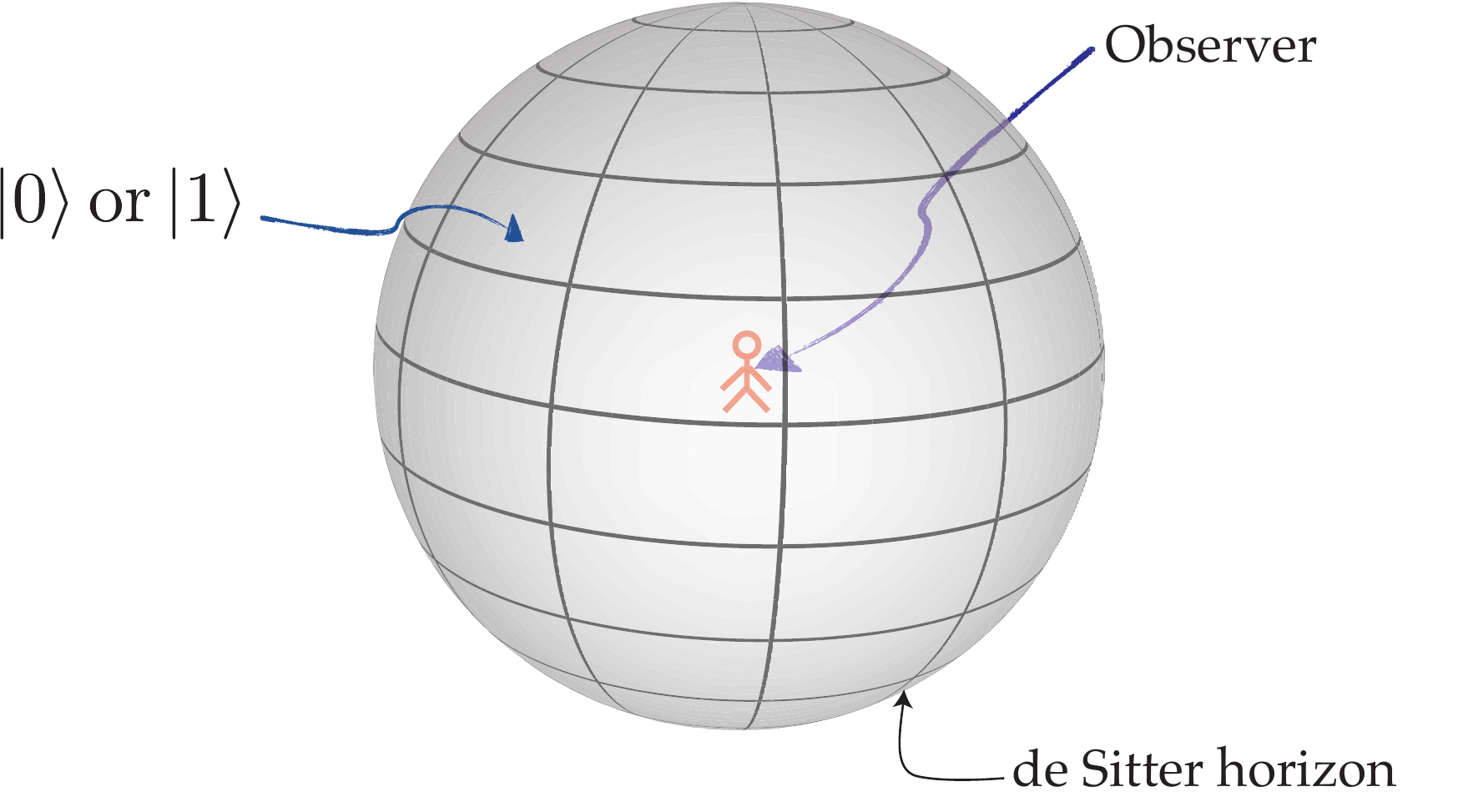} 
\end{center}
{\footnotesize{{\bf Figure 4: }Schematic view of the de Sitter horizon of an observer. Each pixel has area $\ell^2_P$ and can carry either 0 or 1 bit of information.}}
\end{figure*}
From the above equation, we notice that, in addition to terms growing with the area of the horizon, there is a contribution that grows like the total number of e-folds, similar to the one found in \cite{Maldacena:2012xp}. 
\subsection{Bit-per-area prescription}
An alternative and popular way of counting micro-states, borrowed from information theory, is through the bit-per-area prescription \cite{'tHooft:1993gx}. In this prescription, individual pixels are to carry only one bit of Boolean information (either 1 or 0),  so on total there are $\Omega =2^{\np}$ bits of information on the horizon, where again $\np$ is the number of pixels (See Fig.~4).  The entropy in this prescription is defined as the logarithm in base 2 of the total number of information bits, \ie   
\be
S=\log_2 \Omega=\np=
C \(1\over H^2 a_{\rm min}^2\)
\, , 
\ee
where $C$ depends on the coarse graining procedure.  We can also obtain this expression using the density matrix formalism, as explained in Appendix \ref{DM} . Now, using the relationship between the UV cutoff and the lattice spacing; \eq{cutoff}, we obtain
\be
S=  {C\over 2} \(\frac{\Lambda}{H}\)^2\, . 
\label{eq:10}
\ee

The above entropy expression makes perfect sense, since entropy increases as the number of degrees of freedom below the cutoff scale \cite{KeskiVakkuri:2003vj}. This can also be understood as the number of modes that can be fitted on the horizon area which is expected to be dominated by modes with the shortest possible wavelength \ie $\lambda_{\rm min}\simeq {1/k_{\rm max}}\simeq  {1 / \Lambda}$. Hence, the number of modes reads $n_{\rm modes}\sim{(\Lambda/H)^2}$.  This is also consistent with another interpretation of de Sitter entropy; {\em entanglement entropy} which, according to Srednicki's classic calculation, is proportional to the boundary area\footnote{Note that Srednicki's computation was done in a Minkowski background, however, one can extrapolate it to de Sitter space as long as it is deep enough inside the horizon. The boundary area is then $A_{\rm boundary}\sim c \, H^{-2}$, where $0<c<1$. Since entropy is dimensionless, one need to introduce a mass scale to compensate for correct dimensions. Applying the principle of maximum entropy yields the correct scaling $S \sim (\Lambda/H)^2$ .}.

\subsection{The entropy current of the inflaton fluid}
\label{sec:pf}
In the previous subsections, we estimated the EFT cutoff $\Lambda$ by exploiting the fact that entropy scales as $\Lambda^2$, using two different prescriptions. We found that in the decoupling limit, $\Lambda\sim\Delta\phi$. In the following, exploiting the same entropy scaling with the cutoff, we will derive the entropy of the inflaton by considering it as a perfect fluid. We will work in the decoupling limit, keeping $\dot{H}$ only when we consider entropy (area) variation. This limit turns out to be consistent thanks to the slow-roll approximation,  as we will illustrate shortly. We know that in the limit of exact de Sitter, entropy is given by \eq{dsentropy} and is constant, in agreement with the second law of thermodynamics. Now, introducing a slowly rolling scalar field to set the de Sitter phase corresponds to introducing an explicit time-dependence for the Hubble rate. Hence the entropy is no more constant, instead it will increase monotonically with time, again in agreement with the second law. This increase of entropy  according to the discussion of Section (\ref{sec:1}) is due to the increase of the horizon area, which mean that more short wavelength modes can be packed on the horizon. Using the fact that the de Sitter phase is driven by the inflaton, one can therefore write that the dS entropy is just the entropy of the inflaton. the sum of pure de Sitter entropy. This leads to the observation that the increase of entropy is caused by the rolling inflaton, \ie $\dot S=\dot S_\phi$. Moreover, since we are working in the slow-roll limit, we expect that $\dot S_\phi\sim \epsilon$. In fact, in the slow-roll limit, differentiating \eq{dsentropy} with respect to time one gets 
\be
\dot S_\phi=\frac{8 \pi^2}{H}\frac{\dot\phi^2}{H^2}\, , 
\label{dots}
\ee
which is indeed of ${\cal O}(\epsilon)$. 
Let us now construct the inflaton entropy in the perfect fluid picture and see whether it coincides with the expectation \eq{dots}. The energy momentum tensor of the inflaton take the form  
\be
T_{\mu\nu}= p g_{\mu\nu}+ (\rho+p) u_\mu u_\nu\,,  
\ee
where $\rho$ and $p$ are the energy density and pressure of the fluid given by 
\bea
\rho=-\frac12 (\partial\phi)^2 +V(\phi)\quad\textrm{and ~~} p=-\frac12 (\partial\phi)^2 -V(\phi)\, .
\eea
This is the conventional form of a dissipationless perfect fluid energy momentum tensor. The inflaton (and fluid) 4-velocity vector is $u^\mu =\partial^\mu\phi/\sqrt{-(\partial\phi)^2}$ and is normalized such that $u^\mu u_\mu=-1$.  The entropy density of the inflaton can be constructed as usual \cite{weinberg,Weinberg:2008zzc}
\be
s_\phi=\frac{\rho+p}{T}
\label{eq:2}
\ee
where $T$ is the temperature felt by an observer moving with the fluid. By temperature, it is really meant the frequency of the collected photons. In terms of the 4-momentum of the photon $p^\mu$,  formally it is defined as $T=-u_\mu\, p^\mu$. So the entropy is intrinsically observer dependent; it depends both on the coordinate system and the velocity of the observer. However, this is in agreement with the standard interpretation that entropy quantifies the amount of ignorance of the observer about the system.  

In exact dS, entropy is a conserved charge and as any conserved charge, by the Noether theorem, this is a consequence of the conservation of an entropy current $\nabla_\mu \, J^0_S{}^\mu=0$, where the superscript "0" stands for the adiabatic evolution $\dot{H}=0$.  Entropy is then the time component of the integral over space of the entropy current 
\be 
S_\phi=\int\d^3 x\sqrt{^{3}g} \;  J^0_{S}\,,
\label{entropy}
\ee
where  $^{3}g$ is the three-dimensional space metric. We expect that definition to be valid in the decoupling limit. Following the standard procedure \cite{weinberg,Weinberg:2008zzc}, one can construct a current associated with a conserved charge, in our case the entropy. It reads 
\be
J_S^\mu = s_\phi \, u^\mu \, , 
\label{eq:1}
\ee
where $s_\phi$ is the entropy density defined previously. Replacing \eq{eq:1} and \eq{eq:2} in \eq{entropy}, the resulting entropy reads
\be
S_\phi=-\int\d^3 x\sqrt{^{3}g} ~~  {(\partial\phi)^2 \over T}\,  u^0\, .
\ee

In the following, we will compute the entropy in both dS coordinate systems, as well as its increase during a slow-roll phase. In the flat-slicing coordinate system 
\be 
\d s^2=-\d t^2+ \ex{2Ht}\(\d r^2+ r^2 \d\Omega^2 \)\, , 
\label{ds}
\ee
the inflaton pressure and energy density are obtained by considering the homogeneous part of the inflaton $\phi(t)$ as 
\bea
\rho=\frac12 \dot\phi^2 +V(\phi)\quad\textrm{and ~~} p=\frac12 \dot\phi^2 -V(\phi)\, .
\label{rho}
\eea
Likewise, the 4-velocity is simply  $u^\mu=(1,0,0, 0)$. 
Replacing with \eq{rho}, one gets 
\be
S_\phi=\int\d^3 x\sqrt{^{3}g} \;  {2 \pi\over H} \dot\phi^2\,. 
\ee
Let us compute the time derivative of this expression as a check that it represents really the inflaton entropy. In doing so, one has to keep in mind that the integral should be performed on the horizon volume, \ie
\be
S_\phi=\int_{{\rm Vol}_H}\d^3 x\sqrt{^{3}g} \;  J^0_S \equiv \int_0^{1/a H}4 \pi r^2 \d r \, a^3 \;  J^0_S\, .
\ee 
Taking the derivative with respect to time, it is straightforward to get at first order in slow-roll 
\be
\dot S_\phi= {2 \pi\over H}\int_{{\rm Vol}_H}\d^3 x\sqrt{^{3}g} \; 3 H \dot\phi^2 = {8 \pi^2 \over H}{\dot\phi^2 \over H^2}\,, 
\ee
exactly the same as \eq{dots}. 
Although the result obtained above is satisfactory, it was derived in the standard de Sitter metric \eq{ds}.  However, strictly speaking, one should use the de Sitter metric in its static form (see \eg \cite{Frolov:2002va})
\be
\d s^2=-(1-H^2R^2)\d\tau^2+ \frac{\d R^2}{1-H^2R^2} + R^2 \d \Omega^2\, ,
\label{static}
\ee
where the scalar fields complementarity makes sense. Using the Stokes theorem, one can integrate the conservation equation for the entropy current over the horizon volume to get 
\be
{\partial S_\phi\over \partial \tau}= -\int \d^3 x\sqrt{{}^3 g} \;\vec\nabla \cdot \vec{J}_S= -\int\d^2 \Sigma\;  \vec{J}_S\cdot \vec{n} =-4 \pi H^{-2}\, s_\phi \, (\vec{u}\cdot\vec{n})
\label{eq:25}
\ee
where $\vec{n}=(-\sqrt{1-H^2 R^2}, \, 0, \, 0)$ is the unit vector normal to the horizon boundary surface $\Sigma$. Note that $\vec{n}$ is inward-pointing as required by the Stokes theorem and the scalar product is understood to be taken on the horizon. In the static patch the 4-velocity take the explicit form\footnote{Since the surface of constant $\phi$ is spacelike, the overall orientation of $u^\mu$ is chosen such that it is future-oriented. This means that 
$$
u^\mu=\left\{\begin{tabular}{cc} $-\trac{\partial^\mu\phi}{\sqrt{-(\partial\phi)^2}}$ & for $\phi_{,\tau}>0$\\
$+\trac{\partial^\mu\phi}{\sqrt{-(\partial\phi)^2}}$ & for $\phi_{,\tau}<0$
\end{tabular}\right.
$$
} 
\be 
u_\mu=\pm{(\phi_{,\tau}, \phi_{,R}, \, 0, \, 0)} 
\cdot  {H R \over |\phi_{,R}| (1-H^2 R^2)}\, , 
\ee
where we used the usual shorthand $\phi_{,\tau}\equiv\partial\phi/\partial\tau$ and $\phi_{,R}\equiv\partial\phi/\partial R$ and the upper   sign corresponds to $\phi_{,\tau}<0$ and vice-versa. It follows that the inflaton entropy in the static patch is\footnote{As explained in \S\ref{sec:pf}, the temperature measured in the fluid rest frame is $T=-u_\mu p^\mu$. Since there exists a timelike Killing vector $\xi^\mu=(\partial_\tau)^\mu=\delta^\mu_\tau$ in the static coordinates system, one can define a conserved energy associated with the motion along the  time-like Killing vector $E=-\xi^\mu u_\mu=T_{dS}$. It follows (See \eg Section 6.3 of \cite{carroll}) that $T=E/\sqrt{-\xi^\mu\xi_\mu}=T_{dS}/\sqrt{1-H^2 R^2}$. }
\be
s_\phi= -{(\partial \phi)^2 \over T(R)}
\ee
Plugging this expression in \eq{eq:25}, one gets
\be 
{\partial S_\phi \over \partial \tau}=\mp \mbox{sgn}\[\phi_{,R}(\tau, \, R_H)\]\, {8 \pi^2 \over H^3}  \( 1- H^2 R^2  \over H^2 R^2\) \phi^2_{,R}
\label{eq:24}
\ee
Next, let us determine the sign of $\phi_{,R}$. Using the fact that $\phi_{,R}=\[HR/ (1-H^2 R^2)\]\phi_{,\tau}$, one gets 
\be 
{\partial S_\phi \over \partial \tau}={8 \pi^2 \over H^3} 
\phi^2_{,\tau}\, . 
\label{eq:34}
\ee
This means that entropy variation in the static patch is positive, in agreement with the second law. Moreover, by noticing that $\partial_\tau=\partial_t$, expression \eq{eq:34} nicely matches the expected result \eq{dots}, which was calculated in the flat-slicing coordinates.

We are now ready to compute the entropy increase during slow-roll. Again at first order in slow roll, integrating \eq{dots}, we get
\be 
\Delta S=\int_{t_i}^{t_f} {\d S_\phi \over \d t} \, \d t= \frac{8 \pi^2}{H^3}\int_{t_i}^{t_f}\d t  \[ {\d\over \d t}\(\dot\phi\phi\) - \ddot\phi \phi\]\, , 
\ee
where we discarded terms higher than ${\cal O}(\epsilon)$. Now we can successively bound this expression as follows
\be 
 \Delta S<   \frac{8 \pi^2}{H^3} \, \[\, \left|\int_{t_i}^{t_f}\d t   {\d\over \d t}\(\dot\phi\phi\)\right|  + \, \left| \int_{t_i}^{t_f} \d t\,  \ddot\phi \phi\right|\,  \]\, . 
\ee
The first term is integrated directly, while the second one is further bounded using the slow-roll condition $\ddot\phi<3 H \dot\phi$, to give 
\be
 \Delta S<  \frac{8 \pi^2}{H^3} \,    \[| \phi_f \dot\phi_f- \phi_i \dot\phi_i| + {3 H\over 2}| \phi_f^2-\phi_i^2|\]
\ee
Now using the inflaton equation of motion and that \cite{Boubekeur:2012xn} $\sqrt{2\epsilon_\star}\le \Delta\phi/(M_P N)$, we get
\be
\Delta S_{\rm max}\simeq\frac{8\pi^2}{H^2}\[\frac32 \Delta\phi\( 
\phi_i + \phi_f\)+ \sqrt{2}M_P \phi_f\]
\ee
where we neglected an ${\cal O}(1/N)$ term. Note here that we are implicitly assuming that $\epsilon$ is monotonically growing with time. This situation is typical in single-field models either large or small fields. Now using the fact that in both cases, $\Delta\phi/M_P\gg 1$ and that $\phi_{i,f}\gg M_P$, one obtains that 
\be
\Delta S_{\rm max}\simeq12\pi^2\(\Delta\phi\over H\)^2\,,  
\label{smax}
\ee
which by virtue of \eq{cutoff} and \eq{amin}, is in perfect agreement with \eq{eq:10} provided that $C=3$. 

The above result can be easily generalized to the case of multiple scalar fields driving the dS phase. In this case,  the inflationary direction will be a linear combination of these fields. This is especially relevant for multi-field inflation models aiming to avoid super-Planckian excursions by postulating the existence of ${\cal N}>1$ scalar fields with sub-Planckian excursions, while the overall excursion along the inflationary valley is super-Planckian.   The resulting entropy is the sum of individual entropies, owing to the additive property of entropy. From \eq{smax}, we can write   
\be
\Delta S_{\rm max}\simeq12\pi^2\sum_{a=1}^{\cal N}\(\Delta\phi_a\over H\)^2 = 12 \pi^2 {\cal N}\(\overline{\Delta\phi}\over H\)^2\,.
\ee
where $\overline{\Delta\phi}$ is the r.m.s. of all $\phi_a$.  
Using the fact that maximum entropy is bound by the Gibbons-Hawking  entropy, we get  
\be 
\overline{\Delta\phi}\lesssim \frac{M_P}{\sqrt{\cal N}}
\ee
This means again that the total excursion cannot be super-Planckian without invalidating EFT, even if there are many scalar fields, in agreement with the arguments of \cite{Dvali:2007hz}, that is in the presence of ${\cal N}$ degrees of freedom, the cutoff gets rescaled by a factor $1/\sqrt{\cal N}$. In the case of a single inflaton, this means that our main assumption to derive \eq{smax}; $\Delta\phi\gg M_P$ is not valid, \ie super-Planckian excursions contradict entropy bounds \cite{Conlon:2012tz}.

\section{Relationship to Wald entropy}
\label{sec6}
As any conserved quantity, entropy can be defined as a Noether charge \cite{Wald:1993nt}. Wald entropy is defined in this way considering the action as a functional of not only the metric, but also of $R_{\mu\nu\rho\sigma}$ and  its derivatives as well. Starting from a gravitational action $S=\int\d^4x\sqrt{-g} \, {\cal L}(g, R_{\mu\nu\rho\sigma}, \nabla_{\alpha} R_{\mu\nu\rho\sigma}\, , \cdots)$, Wald defines the entropy through the formula \cite{Wald:1993nt}
\be
S_{\rm Wald}=2 \pi \int \frac{\delta {\cal L}}{\delta R_{\mu\nu\rho\sigma}}\; \hat\epsilon_{\mu\nu}  \hat\epsilon_{\rho\sigma}\, \sqrt{h}\,  \d^2\Omega
\label{wf}
\ee
where $\hat\epsilon_{\mu\nu}=-\hat\epsilon_{\nu\mu}$ is the binormal to the horizon surface (bifurcation surface), normalized such that $\hat\epsilon_{\mu\nu}\hat\epsilon^{\mu\nu}=-2$ and  $\sqrt{h}\, \d^2\Omega\equiv\d^2\Sigma$ stands for the induced surface element on the horizon. This formula has been checked in a variety of examples especially black holes (see \eg \cite{Dabholkar:2012zz} for a pedagogical account) and even claimed to be correct for de Sitter \cite{Brustein:2007jj}.

In order to apply this to de Sitter space, let us consider the most general Lagrangian of a real scalar coupled to gravity. In addition, we expect the inflaton to be also coupled to light degrees of freedom to ensure successful reheating . The Lagrangian therefore reads   
\be
{\cal L}={M_P^2\over 2} R-\frac\xi2 R \phi^2 -\frac12 (\partial\phi)^2-V(\phi)+{\cal L}_{\rm reheating}\,.
\label{gl}
\ee
In the above Lagrangian, we also included a non-minimal coupling $\xi$ for generality\footnote{Although the non-minimal coupling is allowed by all symmetries, for simplicity it is usually ignored in most inflation model-building. See However \cite{Okada:2010jf, Kallosh:2010ug}.
}. Although it can always reabsorbed through a Weyl rescaling,  such non-minimal coupling will always be generated through quantum corrections as we explain now. For successful reheating, the Lagrangian ${\cal L}_{\rm reheating}$ {\em should} include couplings of the inflaton to light degrees of freedom. For instance, in addition to a quartic coupling $\lambda\phi^2/4!$ contained in the potential, the inflaton will generically have interaction terms like $y_\psi \phi\bar\psi\psi+ \lambda_\chi \chi^2\phi^2 +\cdots$, where $\psi$ and $\chi$ stand for a generic standard model fermion and boson respectively.   The RGE equation of the non-minimal coupling $\xi$  reads\footnote{Negative sign contributions to $\beta_\xi$ are also possible  if the inflaton is charged under some gauge group.} \cite{buch}
\bea
\beta_\xi=\frac{\(\xi-\frac16\)}{(4\pi)^2}
\[\lambda+\lambda_\chi+4 y^2_\psi +\cdots\]\, , 
\label{RGE}
\eea
will always generate non-vanishing $\xi$. Unless $\xi=1/6$, which is an infrared fixed point, a non-vanishing $\xi$ coupling will be generated through running. It can take any value, however for consistency, the value of the non-minimal coupling is restricted as follows.  
\begin{itemize}
\item Assuming a de Sitter background, the validity of semi-classical methods implies a theoretical upper bound on the non-minimal coupling 
\be
\xi\ll\frac23 S_{dS} \,\({M_P\over\phi_c}\)^2\,.
\ee
This inequality can be also written as an upper bound on the quantity $\xi\phi_c^2/M^2_P$
\be
{\xi\phi_c^2\over M^2_P}\ll S_{dS}
\ee 

\item Conservatively, we also demand that the non-minimal coupling {\it alone} should not overclose the universe during inflation. This implies that the inflaton and thus its excursion should satisfy
\be
\Delta\phi<\phi_c<M_P/\sqrt{\xi}\,.
\label{overc}
\ee
\end{itemize}

Other model-dependent restrictions can be imposed. For instance,  in the case of the Higgs boson, an upper bound \cite{Atkins:2012yn} on $\xi$ can be derived using the LHC data $\xi_H<2.6\times 10^{15}$. 

Applying the Wald formula \eq{wf} to the general inflaton Lagrangian \eq{gl}, one gets 
\be
S_\phi=8 \pi^2 \, {M_P^2\over H^2}\(1-\xi{\phi_c^2\over M_P^2}\)\,.
\label{eq:wald}
\ee
In the above expression, condition \eq{overc} guarantees that the term in parenthesis is always smaller, or at most equal to one. Hence, Wald entropy never overwhelms the Gibbons-Hawking entropy \eq{dsentropy}. Moreover, \eq{eq:wald} means that for entropy to be positive definite and $\xi={\cal O}(1)$, the inflaton excursion should not exceed $M_P$. Moreover, using the results of \S\ref{sec:4}; that entropy scales as $\(\Lambda/H\)^2$, we can read off the cutoff,  
\be
\Lambda=M_P\cdot\sqrt{1-\xi{\phi_c^2\over M_P^2}}\, , 
\ee
where again condition \eq{overc} should be satisfied for consistency. It is also interesting to note that, up to ${\cal O}(1)$ factors, the  above cutoff can also be obtained by considering the $2\to 2$ perturbative unitarity. See Appendix \ref{app:unit} for details. 
\section{Discussion}
\label{sec7}
The problem discussed in this paper is not without reminding us with what was called {\em "SUSY's dirty little secret"} in \cite{Harnik:2004yp}; the problem of fast proton decay through {\em Planck-suppressed} dimension-five operators \cite{Weinberg:1981wj}. The root of this problem is that, for generic ${\cal O}(1)$ dimensionless couplings,  the Planck-scale suppression of these operators is not enough to comply with experimental constraints on proton stability\footnote{See also \cite{Dine:2013nga} for a recent treatment of this problem.}. A similar problem afflicts the Peccei-Quinn solution of the strong CP problem; higher-order operators renders this elegant solution unviable, unless a tremendous fine-tuning is invoked \cite{Kamionkowski:1992mf,Barr:1992qq, Holman:1992us}. Likewise, in inflation, higher-order Planck-suppressed operators, with generic ${\cal O}(1)$ dimensionless couplings inevitably spoil the flatness of the inflationary plateau, putting in question the remarkable success of the  inflationary paradigm. In this work, we revisited this old issue afflicting inflationary model-building, by determining consistently the scale by which the non-renormalizable operators are suppressed in a consistent EFT. We have found, that in the absence of gravity (decoupling limit),  the total excursion of the inflaton $\Delta\phi$ can play the r\^ole of the UV cutoff, as it is the breaking scale of shift symmetry. On the  other hand, once gravity is restored, perturbative unitarity will break down at the Planck scale. This simply implies that the real UV cutoff is the Planck and as a consequence the inflaton excursion cannot exceed this value. If on the contrary, the inflaton excursion exceeds the Planck scale, we are confronted with inconsistencies that manifest themselves in a variety of ways, that go  beyond the well-known $\eta$-problem.   

We argued that since the shift symmetry is broken during inflation, a consistent EFT below the cutoff must include all possible terms that break shift symmetry regardless of their renormalization properties. Then, unitarity and perturbativity (See \eq{eq:20} and \eq{ineq}) constrain the excursions to be sub-Planckian for the theory to make sense. This leads to the conclusion that, in the absence of a UV completion, scenarios with super-Planckian excursions appear to be ill-defined quantum-mechanically. The second manifestation of the inconsistency has to do with entropy; super-Planckian excursions violate de Sitter entropy bounds. We have computed the inflaton entropy using three well-known prescriptions  (bit-per-area, coarse-grained and a standard thermodynamics) and checked that, in the decoupling limit, entropy is proportional to $\Delta\phi^2/M_P^2$, violating the de Sitter entropy bounds for $\Delta\phi>M_P$. We have also considered Wald's geometric definition  of inflaton entropy. We argued, that because of the ubiquitous non-minimal coupling, and in order for this  entropy definition to make sense, excursions have to be sub-Planckian. This conclusion extends also to the multi-field case where there is $\cal N$ inflatons, \ie $\sum_{i=1}^{\cal N} \Delta\phi_i^2\lesssim M_P^2$.

These inconsistencies would not be worrisome if it were not for the observational results, in particular the latest {\sl Planck 2013}  \cite{Planck}, constraining simple inflationary scenarios.  In particular, the non-observation of $B$-modes disfavors the simplest models of chaotic inflation\footnote{The simplest chaotic scenario based on the quartic potential $\lambda\phi^4$ is already ruled-out by the current data, while the one based on the quadratic potential lies outside of the two-standard deviations allowed region \cite{Planck}. }. In this regard, it is really tempting to think that, after all, this is just the harsh experimental verdict on models which, despite their intrinsic simplicity and attractiveness,  are difficult to understand theoretically\footnote{Further criticisms, especially the issue of initial conditions, have been discussed recently in \cite{Ijjas:2013vea}. An additional theoretical objection concerns the cosmological moduli problem, which for $m_{\rm moduli}<H_*$ requires that $H_\star\lesssim 10^7~\gev$, in sharp disagreement with chaotic models \cite{Randall:1994fr, Linde:1996cx}.}.

On the other hand, even scenarios with suppressed tensors, \ie hilltop models \cite{hilltop,Boubekeur:2012xn},  are not free from this theoretical inconsistency\footnote{Notice though that the problem in this case is partially addressed by arguing that non-renormalizable operators like \eq{nr} can generate a local maximum (\ie a hilltop) from which inflation is likely to start  \cite{hilltop}.}. Indeed, a wide class of such scenarios have super-Planckian excursions \cite{Boubekeur:2012xn} which is equally problematic. Therefore, even if in the near future, observational results will eventually single-out one of these scenarios, for instance by measuring accurately the tensor-to-scalar ratio, non-Gaussianity and isocurvature perturbations, one has to address this problem. This seems to be a crucial step before confirming inflation as {\it the} theory for early universe cosmology.

\section*{Acknowledgements}
I would like to thank Paolo Creminelli for invaluable comments on a previous version of the draft. I would like also to thank the ICTP-SAIFR in S\~ao Paulo, and the CERN theory division for hospitality during the completion of this work.  
%\newpage
\section*{Appendices}
\appendix

\section{Details of coarse grained entropy}
\label{cg}
Let us start with the usual Friedmann equations
\be
H^2\simeq{V(\phi)}/{3 M_P^2}\quad \textrm{and~~ }
3 H\dot\phi \simeq-V^\prime(\phi)\, .
\ee
Using the standard definition of the first slow-roll parameter, $\epsilon\equiv {1\over2}M_P^2\(V'/V\)^2$, these equations can be put together, in the simple form
\be
\dot\phi^2 = 2 \epsilon M_P^2 H^2 .
\label{dotphi}
\ee

Now, let us consider the discrete version of the above equation on the horizon surface, which normal vector is $\vec{R_H(\theta,\phi)}\equiv(R=R_H, \theta, \phi )$ in polar coordinates. Instead of $\theta$ and $\phi$, we will use a flat cartesian coordinate $(x, y)$ that we will discretize as $(x, y) \to (i\,a_{\rm min}, j\,a_{\rm min})$, with $i, j=1, 2, \cdots \sqrt{\np}$, where $\np$ is the number of pixels defined in the main text. With this prescription, Eq.~({\ref{dotphi}) becomes a set of $n$ discrete equations

\be 
|\dot\phi_{i, j}(t)| = \sqrt{2 \epsilon_{i,j}(t)} \, M_P H.
\label{eq}
\ee
where the slow-roll parameters are given by
\be
\epsilon_{i,j}=\frac{M_P^2}{8 V(\phi_{i, j})^2} \[\frac{V(\phi_{i+1, j}) - V(\phi_{i, j})}{\phi_{i+1, j}- \phi_{i, j} } + \frac{V(\phi_{i, j+1}) - V(\phi_{i, j})}{\phi_{i, j+1}- \phi_{i, j} }\]^2 \, , 
\label{eps}
\ee
where the explicit time-dependence of $\phi_{i,j}$ and $\epsilon_{i,j}$ is understood.

Equation (\ref{eq}) can be integrated iteratively in time as follows. Given an inflationary potential $V(\phi)$ and a set of $n$ initial conditions $\phi_{i,j}(t_i)$, one can compute the slow-roll parameters $\epsilon_{i,j}(t_i)$, using Eq.~(\ref{eps}), then plug them into Eq.~(\ref{eq}) and integrate between $t_i$ and $t_i+\Delta t$ to obtain the classical solution    
\be
\phi^c_{i,j}(t_i+\Delta t)=\phi_{i,j}(t_i) -\mbox{sgn}\[V^\prime(\phi_{i,j}(t_i))\] \sqrt{2\epsilon_{i,j}(t_i)} M_P H \Delta t. 
\label{sol}
\ee
where $\mbox{sgn}(x)\equiv x/|x|$ is the sign function. Then, one can plug again the above solution in the potential and compute again the slow-roll parameters at $t_i+\Delta t$ and turn the crank gain, until reaching $t_f$. Therefore all we need is  $\np$ initial conditions on the $\np$ pixels. However, since we can swap the pixels without changing the outcome of the final (observed) state, the total number of micro-states is therefore $\Omega=\np!$. 

Up to now, we neglected the quantum fluctuations of the inflation. Indeed, each Hubble time, the inflaton gets kicked randomly up or down the slope of the inflationary potential by the amount $\delta\phi_q\simeq \(H/2 \pi\)$. Thus, to the classical solution \eq{sol}, one has to superimpose $\delta \phi_q$ each e-fold. This has the effect to increase the number of possible initial conditions by a certain factor that we will determine now. We can still solve \eq{eq} iteratively by including the quantum kicks each e-fold. To do so, consider that we discretize time into $N$ e-folds, each one has $\Delta t=H^{-1}$. The  complete solution after one e-fold will be  
\be 
\phi_{i,j}(t_i+H^{-1})=\phi^c_{i,j}(t_i+H^{-1})\pm \frac{H}{2 \pi} \, ,  
\label{sol1}
\ee
which corresponds to more possible initial conditions as anticipated.  Indeed, for a given classical initial condition (ignoring $\delta\phi_q$) $\phi^{c}_{i,j}(t_i)$, there will corresponds 2 complete solutions after 1 e-fold $\phi^{c}_{i,j}(t_i) \pm \delta\phi_q$, taking into account quantum kicks. After 2 e-folds, there will be more corresponding initial conditions; in total 4 initial conditions
\bea
\phi_{i,j}(t_i+2 H^{-1})&=&\phi^c_{i,j}(t_i+2 H^{-1})\\
\phi_{i,j}(t_i+2 H^{-1})&=&\phi^c_{i,j}(t_i+2 H^{-1})\pm  2\, \(\frac{H}{2 \pi}\) \,, 
\eea
which multiplicities are as follows.  The first one corresponds to a kick upward and an other downward the inflationary potential or vice versa, thus in total two corresponding initial conditions. The second and third one correspond to two consecutive kicks either upward or downward, so in total they correspond to an initial condition each. In total we have then 4 initial conditions at 2 e-folds. After $m\le N$ e-folds, 
solutions become
\be
\phi_{i, j}(t_i+ m\, H^{-1})= \phi^c_{i, j}(t_i+ m\, H^{-1})+ (m-p) \frac{H}{2 \pi} - p\, \frac{H}{2 \pi} \, 
\ee
where $p\le m$. It is easy to convince oneself that the number of corresponding initial conditions will be $2^p$ times $n!$. The factor  $2^p$ is easily understood as corresponding to how many ways one can pick $p$ kicks downward the potential out of the available $m$ in the case of positive $V^\prime$, and vice versa for negative $V^\prime$.   Summing over all $N$ e-folds, the total number of initial conditions is therefore 
\be
\Omega= \np!\sum_{p=0}^{N} \frac{N!}{p!(N-p)!}=2^N \, \np!\, .
\ee

\section{Relating the UV cutoff to $a_{\rm min}$ and $\Delta\phi$}

\label{app:amin}
In this appendix we will determine the precise relationship between $\Lambda$ and $a_{\rm min}$. By putting the inflaton on a lattice of spacing $a_{\rm min}$, one gets as usual a natural UV cutoff $\Lambda\propto 1/a_{\rm min}$. The precise coefficient is determined as follows. Consider a massive scalar field in flat space with a standard dispersion relation $E_p^2=\vec{p}{\,}^2+m^2$. Let us first compute the following Green function 
\begin{equation}
\vev{\phi(t, \vec{x}+\vec{r})\phi(t, \vec{x})}=\int\frac{d^3p}{(2\pi)^3 }{1 \over 2 E_p} e^{i \vec{p}\cdot\vec{r}}\, , 
\label{eq:green}
\end{equation}
and then take the limit of small spatial separations. The integral in the Green function can be performed exactly to yield the standard expression  
\begin{equation}
\vev{\phi(t, \vec{x}+\vec{r})\phi(t, \vec{x})}={m\over 4 \pi^2 |\vec{r}|} K_1(m|\vec{r}|)\,, 
\end{equation}
where $K_n(z)$ stands for modified Bessel functions of the second kind. Expanding this expression for small spacial separation, and making the identification $|\vec{r}|\to a_{\rm min}$ one gets 
\begin{equation}
\vev{\phi(t, \vec{x}+\vec{r})\phi(t, \vec{x})}\xrightarrow[r\to a_{\rm min}]{}\frac1{4 \pi^2 a^2_{\rm min}}
\label{vev}
\end{equation}
On the other hand, taking the limit $|\vec{r}|\to 0$, one gets  
\begin{equation}
\vev{\phi(t, \vec{x}+\vec{r})\phi(t, \vec{x})}=\int\frac{d^3p}{(2\pi)^3 }{1 \over 2 E_p} e^{i \vec{p}\cdot\vec{r}}\xrightarrow[|\vec{r}|\to 0]{}\frac{\Lambda^2}{8\pi^2}
\label{eq:vev}
\end{equation}
where we have introduced a momentum cutoff $\Lambda$ and neglected $m$. Comparing these expressions, we get 
\be
a_{\rm min}=\frac{\sqrt{2}}{\Lambda}\,. 
\label{cutoff}
\ee
Now, one can also write the Green function \eq{eq:green} for small $|\vec{r}|$ as  
\begin{equation}
\vev{\phi(t, \vec{x}+\vec{r})\phi(t, \vec{x})}\xrightarrow[r\to 0]{}\vev{\phi(\vec{x})^2}\gtrsim\Delta\phi^2
\label{eq:exc}
\end{equation}
where  $\Delta\phi\equiv|\phi_f-\phi_i|$  is the total inflaton excursion.  Therefore, combining Eqs (\ref{vev}) and  (\ref{eq:vev}) with \eq{eq:exc} we finally have 
\be
1/a_{\rm min}\simeq 2 \pi \Delta\phi
\label{amin}
\ee

\section{Bit-per-area density matrix}
\label{DM}
In this appendix, we shall compute the inflaton entropy using the density matrix formalism. We can label the  pixels with a set of orthonormal ket vectors $\ket{\alpha}$, satisfying $\vev{\alpha|\beta}= \delta_{\alpha\beta}$ and $\sum_{\alpha}\ket{\alpha}\bra{\alpha}=\mathbf{1}$, where $\alpha$ run from 1 to $\np$. Each pixel can be either carry 0 or 1 as Boolean information, so the normalized state vector reads
\be
\ket{\alpha}=\frac{1}{\sqrt{2}}\(\ket{0}_\alpha+\ket{1}_\alpha\)
\ee
Then, the horizon state vector reads
\be
\ket{\psi}=\prod_{\alpha}\ket{\alpha}=\frac{1}{2^{\np/2}}\prod_\alpha\(\ket{0}_\alpha+\ket{1}_\alpha\)\, , 
\ee
from which we can construct  the density operator  
\be
\hat\rho=\ket{\psi}\bra{\psi}=
\frac{1}{{2^{\np}}}\prod_{\alpha\beta} \Big{(} \ket{0}_\alpha+\ket{1}_\alpha \Big{)}
\(\bra{0}_\alpha+\bra{1}_\beta\)
\ee
Therefore, the density matrix is simply 
\be
\rho_{i, j}= {1\over 2^{\np}}
\label{dm}
\ee
where $i,\, j$ labels the basis of all possible products $\prod_\alpha\ket{ 0 \textrm{ or } 1}_\alpha$. 
It follows that entropy in this prescription is 
\be 
S=\np
\ee

\section{Unitarity in detail}
\label{app:unit}
In this appendix, we will determine the tree-level unitarity cutoff for the theory given by the Lagrangian \eq{gl}. As usual, the unitarity of $\phi\phi\to\phi\phi$ scattering amplitude demands 
\be
|\textrm{Re}\, a_J|\le\frac12
\ee
where $a_J$ is the $J$-th partial wave amplitude
\be
a_J(s)=\frac{1}{32\pi}\int_{-1}^{1}d(\cos\theta)\, P_n(\cos\theta) {\cal A}(s,\theta)
\ee
and ${\cal A}(s,\theta)$ is the $2\to 2$ scattering amplitude. From the Lagrangian \eq{gl}, and discarding for the moment the potential $V(\phi)$, we get the interactions
\be
{\cal L}=-\frac1{2\overline{M}\!_P} h_{\mu\nu}\del^\mu\phi \del^\nu\phi -\frac{\xi}{2 \overline{M}\!_P}\square h^\mu_\mu\phi^2\, , 
\ee
where we have neglected the mixing and the effective Planck scale\footnote{When linearizing around the FRW background, we should use this scale instead of $M_P$, since it will give the correct normalization of the graviton kinetic term. } ${\overline M}\!_P\equiv M_P\cdot\sqrt{1-\xi{\phi_c^2\over M_P}}$. In the center-of-mass frame, the $s$-channel amplitude reads\footnote{For simplicity, we do not consider the full scattering amplitude, that includes the two remaining channels, because it has the usual infrared divergence associated with the exchange of on-shell massless states in the $t$- and $u$-channel.  Their inclusion does not change qualitatively the conclusion of our discussion. }
\be
{\cal A}(s, \theta)=-\frac{2 s}{3\overline{M}^2_P}\[\(1-6\xi\)^2+ \frac12 (1 -3 \cos^2\theta)\]
\label{amplitude}
\ee
From the amplitude \eq{amplitude}, one gets the $J=0$ ($s$-wave) partial-wave 
\be 
a_{J=0}(s)=\frac{s (1-6 \zeta )^2}{96 \pi {\overline{M}\!_P}^2  }\,. 
\ee
Unitarity of this partial-wave amplitude implies a cutoff 
\be
\Lambda_{J=0}=  {\sqrt{\frac{\pi }{3} (1-\xi \phi_c^2/{M_P^2})}\over {|\xi -1/6|}}\cdot M_P\, .
\ee
For small $\xi$, including vanishing $\xi$,  but still away from the conformal limit \ie $0\le|\xi|< 1/6$, the cutoff reduces to $\sqrt{\frac32}m_P$.  While for larger $|\xi|$ but still satisfying \eq{overc}, \ie $1/6<|\xi|<(M_P/\phi_c)^2$, the cutoff  is $\Lambda_0\simeq m_P \frac{1}{\sqrt{6} |\xi|}$. Since we are assuming  that $|\xi|>1/6$, this scale  is bound as follows $\Lambda_0\lesssim \sqrt{6}m_P$, again parametrically $m_P$. This value in agreement with the findings of \cite{Barbon:2009ya, Hertzberg:2010dc}, in the absence of interactions \ie a potential for $\phi$. Also, it is useful to notice that, as expected in the conformal limit, unitarity of the $J=0$ partial wave extends to arbitrarily high energies. \\ 

On the other hand, the $J=2$ partial wave ($d$-wave) is 
\be 
a_{J=2}={s\over 120 \pi  {\overline{M}\!_P}^2}\, , 
\ee
and unitarity gives the cutoff 
\be 
\Lambda_{J=2}=  2\sqrt{15 {\pi } \(1-\xi \frac{\phi_c^2}{M_P^2}\)}\cdot M_P\, .
\ee
For small $\xi$, the cutoff $\Lambda_{J=2}$ approaches $\sqrt{\frac{15}{2}} m_P$. All the remaining partial waves vanish, so the perturbative unitarity cutoff is just $\Lambda=\min(\Lambda_0, \Lambda_2)$. In general, it is parametrically $m_P$. This is so even in the regime $1/6<|\xi|<M_P^2/\phi_c^2$ where in this case $\Lambda_2\simeq M_P/|\xi|\lesssim 6 M_P$ which is again of ${\cal O} (m_P) $. \\
Now, let us now consider unitarity from interactions coming from the potential $V(\phi)$. To this end, it is convenient to work in the Einstein frame where the potential reads
\be
V_E(\phi)=\frac{V(\phi)}{1-\frac{\xi\phi^2}{M_P^2}}\,,
\ee
As noted in \cite{Barbon:2009ya, Hertzberg:2010dc}, for $\xi \phi^2_c\ll M_P^2$, one can expand the denominator and estimate the unitarity cutoff \eg for $3\to 3$ scattering as $\Lambda\simeq {M_P/ \xi}$,  where we neglected some quartic self-coupling. On the other hand, for large excursions, one cannot expand the denominator, and the resulting unitarity bound is weakened by the ratio $\Delta\phi/M_P$. As a consequence, the cutoff is always parametrically $m_P$.

\end{document}